# Enhancing Rumor Detection in Social Media Using Dynamic Propagation Structures

Shuai Wang[1,2], Qingchao Kong[1,3,4*], Yuqi Wang[1], Lei Wang[1,3,4]
[1]State Key Laboratory of Management and Control for Complex Systems, Institute of Automation, Chinese Academy of Sciences, China
[2]University of Chinese Academy of Sciences, China
[3]Shenzhen Artificial Intelligence and Data Science Institute（Longhua）, China
[4]The State Information Center, Beijing 100045, China
{wangshuai2017, qingchao.kong, yuqi.wang, l.wang}@ia.ac.cn

*Abstract*—Social media, such as Facebook and Twitter, has become one of the most important channels for information dissemination. However, these social media platforms are often misused to spread rumors, which has brought about severe social problems, and consequently, there are urgent needs for automatic rumor detection techniques. Existing work on rumor detection concentrates more on the utilization of textual features, but diffusion structure itself can provide critical propagating information in identifying rumors. Previous works which have considered structural information, only utilize limited propagation structures. Moreover, few related research has considered the dynamic evolution of diffusion structures. To address these issues, in this paper, we propose a Neural Model using Dynamic Propagation Structures (NM-DPS) for rumor detection in social media. Firstly, we propose a partition approach to model the dynamic evolution of propagation structure and then use temporal attention based neural model to learn a representation for the dynamic structure. Finally, we fuse the structure representation and content features into a unified framework for effective rumor detection. Experimental results on two real-world social media datasets demonstrate the salience of dynamic propagation structure information and the effectiveness of our proposed method in capturing the dynamic structure.

*Keywords—rumor detection, dynamic propagation structure, diffusion process*

## I. Introduction

With the rapid development of social media, some services such as Facebook, Twitter, and Sina Weibo, have become important platforms for Web users to share information and express opinions. However, by taking advantage of social media platforms, some misinformation also spread quickly, which has caused severe social problems. Especially at the time of breaking news, people usually desperately search for information updates. However, some of them just tuned out to be rumors posted by users with malicious intentions, for example, to cause public panic or even social unrest.

To debunk rumors on social media, consulting experts or authoritative organizations are not always possible, because rumor information usually spreads very fast and may reach a large number of users before experts or authoritative organizations come to a conclusion. Consequently, there are urgent needs to automatically detect rumors at the early stages of their diffusions so as to minimize the potential negative impact.

Most previous work on rumor detection leverage content information. Because of the self-correcting property of the platform, there are often some negation and question words in the comments of rumors [1]. To extract high-level and latent features automatically, deep learning models have been applied. To identify rumors from content information, Ma et al. [2] allocate posts in a time sequence to extract context features using the Recurrent Neural Network (RNN) model. To improve the performance of the work by Ma et al. [2], CNN-based model [3] is used to exploit the interactive information between reposts (i.e. replies with comments).

Furthermore, propagation structure provides salient information, which reflects the characteristics of diffusion process. The responsive relationship among posts form the propagation structure. The propagation structure is a tree and the root node is the original tweet. Previous methods use structural information in identifying rumors by different means: evaluating the similarity of propagation tree [4], calculating the properties of tree structure [5], and utilizing implicit links such as hashtag and Web linkage [6]. Later improvements make better use of the structural information by combining temporal features of propagation tree [7, 8, 9]. As these methods are all based on the traditional feature-based models, the main drawback is that they could not capture the dynamic interactive information of posts in different time period.

Note that the propagation structure is constantly changing during the diffusion process as shown in Fig. 1, and this dynamic structural information is an effective indicator for distinguishing rumors and non-rumors [10]. Dynamic propagation structures of rumor and non-rumor are usually more distinguishable than those of the final static structures. Fig. 1 shows two dynamic propagation structure examples about rumor and non-rumor. It is difficult to differentiate the two final structures, while the structures evolving processes are distinguishing.

In this paper, we present a Neural Model using Dynamic Propagation Structures (NM-DPS) for rumors detection. To avoid the weaknesses in previous methods, we consider the evolving process of propagation structure over time. We partition the structure into several segments according to posting time of each tweet and then encode each segment into a vector as the input of neural network. Then we use attention based Bi-directional Gate Recurrent Unit (BiGRU) to learn a representation for propagation structure. We propose an temporal attention mechanism to focus on more informative phases during the propagation process. In summary, our model integrates textual content and dynamic propagation structures into a unified framework. The contributions of our work are as follows:

- We propose a new approach to make use of the dynamic propagation structure, which partitions the entire structure into several sub-structures to model the propagation process.
- We propose a fusion model that integrates propagation structure and textual content information into a unified framework for rumor detection.
- Experiments on two real-world social media datasets show the effectiveness of our model as well as the incorporation of dynamic propagation structures in rumor identification and early detection tasks.

---

*Corresponding author: Qingchao Kong (qingchao.kong@ia.ac.cn)

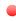

Fig. 1. Examples of evolving processes of propagation structures during the message diffusion. Each node represents a tweet, and edges among the connected tweets represent the relationship of retweeting or reposting. The red root nodes represent the original tweet. The colors of nodes represent the posting time phases. Nodes in the same color are of the same phase. The right gray graphs are the propagation structures at the end of the diffusion. We evaluate the similarity of the propagation structures between rumor and non-rumor in each time units by Weisfeiler-Lehman graph kernel. It shows that the sub-structures during the evolving process are more discriminative than the final structures.

## II. RELATED WORK

There are two categories of rumor detection tasks, namely post-level [11] and event-level [2, 3, 12, 13]. The post-level task is challenging as it only uses content information of the given post and needs much priori knowledge. In this paper, we consider the event-level rumor detection task. There is a set of posts in each event and the objective is to identify whether the event is a rumor by leverage the posts in it. Below we summarize the related work on rumor detection based on the information they utilize.

Most content-based methods leverage the characteristics of posts for rumor detection. As the distribution of negations and question words are distinguishable for rumors and non-rumors, Zhao et al. [1] identify enquiring sentences via pattern matching and judge the correctness of posts by these enquiries. Ma et al. [2] split posts in an event into several groups according to the post time and then they use RNN to capture the context information of the chronological groups. Yu et al. [3] split all the posts into a fixed number of groups, and learn the representations of each group. They then use CNN-based model to extract features. Since the original post contains rich content information, Xu et al. [14] handle the original post and reposts separately at word-level and post-level with Bi-directional Long Short Term Memory (Bi-LSTM).

In addition to content information, the structure of propagation structure also provides evidence for detecting rumors. Inspired by deep walk [15], Wu et al. [4] propose a random walk graph-kernel to model the similarity of propagation trees. The random walk kernel has the drawbacks such as runtime problems and tottering property, which limit the generalization performance of the classifier. Kwon et al. [5] use a Random Forest based method to extract several structural features, such as density, clustering coefficient of the propagation tree and in-degree, out-degree of the node. Sampson et al. [6] improve the performance by adding implicit links as hashtag linkage and Web linkage to enrich the structural information. To exploit the information of dynamic propagation structures, Ma et al. concatenate the structural features in different phases, and use tree-based kernel with time information involved to measure the similarity of nodes [9]. To avoid feature engineering, Ma et al. [12] propose a tree-based Recursive Neural Network (RvNN) to cater the structure of propagation tree. But, they cannot capture the dynamic propagation process.

In this paper, we propose a neural model NM-DPS for rumor detection in social media, which uses dynamic propagation structures and textual content. To capture the dynamic information, we partition the propagation structure according to posting time of each tweet. We learn the representation of the structure with Bi-directional GRU model and introduce temporal attention mechanism to focus on the phases with informative structures. Our model also fuses propagation structure and content information for the rumor detection task. Experimental results on two social media datasets show the effectiveness of our model as well as the incorporation of dynamic propagation structures in rumor detection.

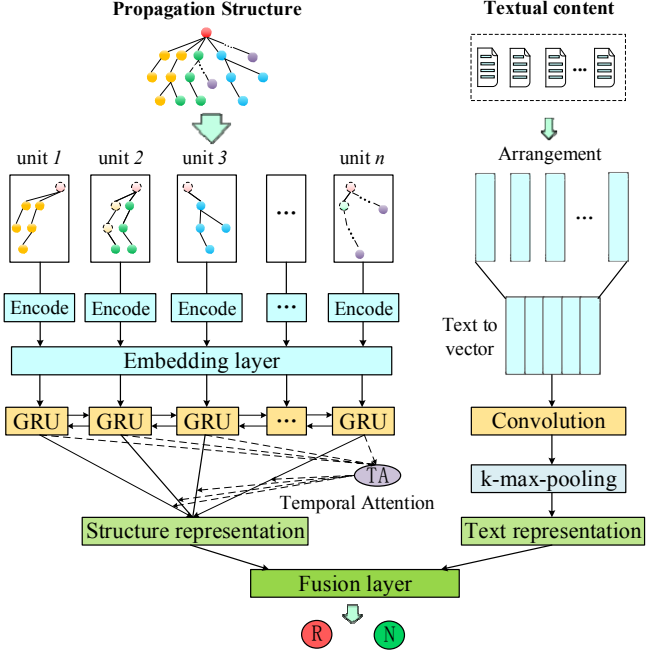

Fig. 2. An overview of our NM-DPS Framework. The inputs are propagation structure and all the texts of the event. We partition the structure into several sub-structures by posting time of each tweet, which is represented by different color. The output is a predictive result that whether the event is a Rumor (R) or Non-rumor (N).

## III. PROPOSED MODEL

### A. Problem Definition

In this paper, the rumor identification task is at event-level. Each event $E_i$ contains an original post $r_i$ and a set of subsequent responsive posts $p_{i*}$ with content information $C_i$, i.e., $E_i = \{r_i, p_{i1}, p_{i2}, \dots, p_{im}\}$, where $m$ denotes the number of responsive posts. In the process of information diffusion, a propagation tree is formed with the reply relations as its edges and the original tweet as its root. The tree structure is evolving over time, resulting in dynamic propagation structures, denoted as $S_i$. The task of rumor detection is to predict whether the event is rumor or not at an early stage, which can be formulated as

$$y_i = f(S_i, C_i) \qquad (1)$$

where $y_i$ is the prediction result that $y_i$ is a rumor or not, $f$ is our NM-DPS model.

In our proposed NM-DPS model, we learn representations for the propagation structures and textual contents by *structure network* and *content network* respectively (see Fig. 2). The *structure network* partitions the propagation structure according to posting time and uses temporal attention mechanism to focus on phases contained informative structures during the diffusion process. *Content network* allocates the contents of posts in a chronological sequence, and learns embedding vectors for each content which contains semantic and sentiment information.

### B. Structure Network

For the propagation structure of each event, we first partition the propagation structure by posting time of each tweet. Specifically, we equally split the entire diffusion time span into several time units, and tweets in each time units form a sub-structure. Then we encode all the sub-structures into vectors as the input of Bi-directional Gate Recurrent Unit (BiGRU). And we also propose the temporal attention mechanism to focus on time units which contain informative structures.

*1)* Propagation structure partitioning. To capture the variation of the dynamic propagation structure during the diffusion process, we partition the structure according to the posting time of each tweet. Specifically, we split the entire time span into several time units with fixed duration. The time units are arranged in a chronological sequence, denoted as $[t_1, t_2, \dots, t_{n_s}]$, where $t_i$ is the *i*-th time unit.

*2)* Structure encoding and embedding. Each time unit contains a sub-structure formed by the tweets posted during this time unit as shown in Fig. 2. We encode each sub-structure and embed it into a dense vector. Note that the representation of structural features can be optimized during the training process. Specifically, we make use of the following three structural features in each time units:

(1) The *ratio of retweets and reposts* in each layer of propagation tree. We denote the ratio of *j*-th layer in the *t*-th time window as $p_{tj}$.
(2) The *ratio of post numbers in adjacent layers*, which reflect the feature of diffusion depth in the spreading process. We denote the ratio of $N_{tj}$ and $N_{(t-1)j}$ as $l_{tj}$, where $N_{tj}$ is the post number of *j*-th layer in *t*-th time unit.
(3) The *number of posts*. We denote the number of *j*-th layer in *t*-th time window as $n_{tj}$.

Then we embed the *t*-th sub-structural information as

$$x_t = embedding(p_t, l_t, n_t) \qquad (2)$$

where $p_t, l_t, n_t$ are lists which are composed of $p_{tj}$, $l_{tj}$, $n_{tj}$ respectively.

*3)* Representation learning for each sub-structure. Since all the time units are arranged in a chronological order, we use RNN to cater the form. For better modeling the context information of the whole evolving process of structures, we adopt BiGRU to learn representations for dynamic structures. The following equations are used for a GRU layer:

$$z_t = \sigma(x_t U_z + h_{t-1} W_z) \qquad (3)$$

$$r_t = \sigma(x_t U_r + h_{t-1} W_r) \qquad (4)$$

$$c_t = \sigma(U_h x_t + W_h(r_t \odot h_{t-1})) \qquad (5)$$

$$h_t = (1 - z_t) \odot h_{t-1} + z_t c_t \qquad (6)$$

where $z_t$ and $r_j$ are update gates and reset gates respectively, $h_t$ is the hidden state at *t*-th time step and $\odot$ is element-wise multiplication, $x_t$ is the input, which is the embedding of sub-structure in *t*-th time unit. *W* and *U* are parameters.

We obtain the hidden state at each time step, which is formulated as

$$[h_1, h_2, \ldots, h_n] = BiGRU([x_1, x_2, \ldots, x_n]) \quad (7)$$

To aggregate the most informative sub-structures, we propose a temporal attention mechanism to focus on the time units which contains salience structure information and suppress the noisy time units:

$$S = TimeAtt([h_1, h_2, \ldots, h_n]) \quad (8)$$

where *S* is the representation of the structure information in the whole evolving process. Specifically, the temporal attention mechanism is computed as follows:

$$u_t = tanh(W_h h_t + b_n) \quad (9)$$

$$\alpha_t = \frac{exp(u_s^T u_t)}{\sum_t exp(u_s^T u_t)} \quad (10)$$

$$S = \sum_t \alpha_t u_t \quad (11)$$

We map the state $h_t$ into a vector representation $u_t$ into a hidden space with a dense layer. Then, an attention weight is calculated by the similarity between $u_t$ and a trainable vector $u_s$, where $u_s$ indicates the informative sub-structures in the hidden space. We normalize the attention weight with a softmax function. The finally structure representation *S* is the weighted summation of the $u_t$.

### C. Textual Content Network

Due to the self-correcting property of crowdsourcing [14], users often support the correct information and question or deny the rumors. To capture the repost content information for detecting rumors, we learn representations for each repost in *content network*, which contain the information of word order and sentiment.

Unlike CAMI [3] which split reposts into several groups, we utilize each repost independently in chronological order. Since there are more interactive relations in post-level than the group-level, we can get richer context information in post-level process. For each event, allocation of its reposts can be formulated as $R_i = [r_{i1}, r_{i2}, \ldots, r_{in}]$, where $r_{ij}$ is the text content of the *j*-th reposts in event $E_i$, *n* is the maximum number of posts in all events.

To learn representations for each repost, we choose the same approach as CAMI [3], which is an unsupervised approach called *paragraph vector* [16]. It is an advanced framework of word2vec [17], and the outputs are the representations of textual content. The learned representation the of all the reposts in an event is denoted as $\hat{X}$. A Convolutional Neural Network (CNN) is used to extract features. Specifically, we apply one-dimensional convolution operation with different size filters $W \in \mathbb{R}^{h*d}$ to extract multiple features. We also use dynamic k-max-pooling [18] to obtain the k-largest value of the feature map, denoted as $g^{k-max}$, which can better capture the relations of long-range element in the sequence. The *j*-th feature map is formulated as

$$g_j^{k-max} = L(ReLU(\langle W_j, \hat{X}\rangle_F + b)) \quad (12)$$

where *L* is k-max pooling operation, $W_j$ is the filter with $h_j$ height, *F* is Frobenius inner product and *b* is bias. The final representation of *textual content network* is calculated by

$$C = \phi_s(g_1^{k-max}, g_2^{k-max}, \ldots, g_m^{k-max}) \quad (13)$$

where $\phi_s$ is the converting and concatenation operation, *m* is the number of filters.

### D. Optimization and Prediction

For the fusion layer, we merge the representations of *content network* and *structure network* by a concatenate operation and a follow-up fully connected layer. Finally, the probability of an event being a rumor with a softmax function:

$$X = \phi(C, S) \quad (14)$$

$$y = \sigma(f(X)) \quad (15)$$

where *y* is the prediction result indicating whether the event is a rumor, *f* is the fully connected layer, $\sigma$ is the sigmoid function. S is the structure representation in eq. (8), C is the representation of textual content network in eq. (13).

We use cross-entropy as loss function, i.e.

$$L = -\sum_{i=1}^{M}(y_i log\hat{y}_i + (1 - y_i)log(1 - \hat{y}_i)) \quad (16)$$

where $y_i$ and $\hat{y}_i$ are and prediction result and true label of the *i*-th event respectively. We adopt Adam [19] as optimizer to speed up the convergence in the training process and dropout [20] strategy in the penultimate layer to avoid overfitting.

## IV. EXPERIMENT

In this section, we first compare our method with several baseline methods for rumor detection and conduct experiments with different detection deadlines (i.e. prediction time). Then we will show the effectiveness of dynamic propagation structures in rumor detection task through an ablation study.

### A. Datasets

For comparison, we use two large microblog datasets collected by Ma et el. [2, 12], namely Sina Weibo and Twitter dataset, which have been used in [2, 3, 7]. Details of the datasets are shown in Table I.

TABLE I. STATISTICS OF THE DATASETS

| Statistic | Weibo | Twitter |
|---|---|---|
| #Posts | 3,805,656 | 25,234 |
| #Events | 4,664 | 1158 |
| #Rumors | 2,313 | 579 |
| #Non-Rumors | 2,351 | 579 |

TABLE II. COMPARATIVE EXPERIMENTAL RESULTS OF RUMOR DETECTION METHODS

| Method | Class | Weibo | | Twitter | |
|---|---|---|---|---|---|
| | | Accuracy | $F_1$ | Accuracy | $F_1$ |
| SVM-TS [8] | R | 0.857* | 0.861* | 0.756 | 0.761 |
| | N | | 0.857* | | 0.751 |
| GRU-2 [2] | R | 0.910* | 0.914* | 0.769 | 0.772 |
| | N | | 0.906* | | 0.764 |
| PPC [7] | R | 0.921* | 0.923* | 0.769 | 0.742 |
| | N | | 0.918* | | 0.794 |
| CAMI [3] | R | 0.933* | 0.933* | 0.797 | 0.783 |
| | N | | 0.932* | | 0.778 |
| **NM-DPS** | R | **0.943** | **0.945** | **0.836** | **0.832** |
| | N | | **0.938** | | **0.839** |

(R: Rumor, N: Non-Rumor, *: the result taken from the corresponding paper)

### B. Experiment Settings

We choose three kinds of convolutional filters with different heights, 5, 6, 7, and set the number of feature map to 30 for each kind of filter. We set the timespan to 20 minutes, word dimension to 50, and the dropout rate to 0.5. Five-fold evaluation is conducted on the above two datasets. Evaluation metrics used in this experiment are accuracy and F-1 to evaluate the performance of the proposed model. Since the Weibo dataset is same as [2, 3, 7], we take their experiment results from the original papers respectively. However, some tweets have been deleted when the Twitter dataset is collected by the given ID, we re-run the methods from related work and obtain the prediction results using the current Twitter dataset.

### C. Baseline methods

There are two different types of comparative methods, i.e. tradition feature-based methods and recently proposed neural-based methods. For the feature-based methods, we choose SVM-TS which also incorporate structural information. For the neural based methods, we choose three recently proposed methods which achieve state of the art results, namely GRU-2 [2], CAMI [3] and PPC [7].

(1) *SVM-TS* [8] constructs a graph kernel to combine the content and propagation structure information and adopts a SVM-based time series model to capture the information in different phases.

(2) *GRU-2* [2] is a RNN based model, which use Bi-directional GRU to capture the temporal features and content information.

(3) *CAMI* [3] is a CNN-based model, which exploit textual features of posts with two convolutional layers.

(4) *PPC* [7] extracts features of propagation path among users in an event by combining CNN and RNN model.

### D. Experimental Results

*1) Rumor Detection Results.*

As shown in Table II，our proposed model NM-DPS achieves substantial gains in all evaluation metrics over the previous state-of-the-art methods on both of the two datasets. Deep learning methods such as GRU-2, CAMI and PPC perform better than traditional feature-based methods SVM-

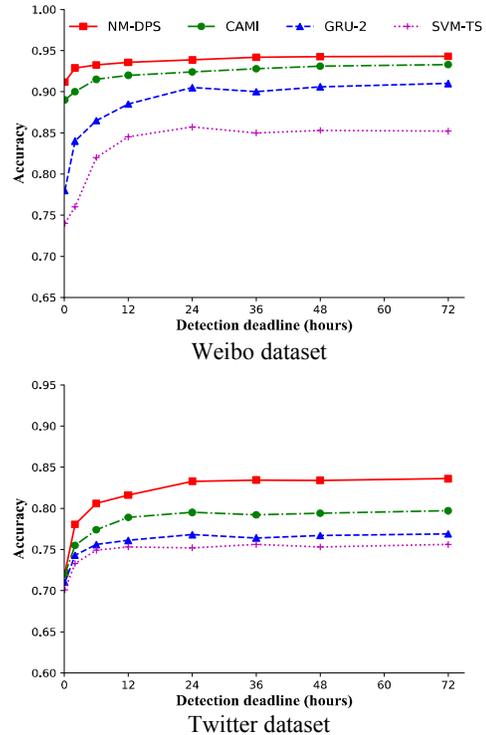

Weibo dataset

Twitter dataset

Fig. 3. Early detection of rumors

TS. One possible reason for this is that deep learning methods can be applied to complicated scenarios and they can automatically extract latent and interactive features.

Table II also shows that the performance gain in Twitter dataset is much larger than that in the Weibo dataset. For example, comparing to CAMI, the accuracy of our proposed model increases by 0.010 in the Weibo dataset, which is less than 0.039 in the Twitter dataset. A possible reason is that, some of tweets in the Twitter dataset are deleted, while the propagation structure remains intact when we collect the data. In this case, textual content based methods, such as GRU-2 and CAMI, do not perform well. The proposed NM-DPS model integrates structural information of reposts and content information, so it outperforms GRU-2 and CAMI.

*2) Ablation Study*

To further analyze the effectiveness of the dynamic propagation structure, we compare the rumor prediction results between the following components:

TABLE III. EFECT OF DIFFERENT COMPONENTS IN RUMOR DETECTION

| Method | Class | Weibo | | Twitter | |
|---|---|---|---|---|---|
| | | Accuracy | $F_1$ | Accuracy | $F_1$ |
| *Static-Struct* | R | 0.557 | 0.550 | 0.561 | 0.526 |
| | N | | 0.563 | | 0.595 |
| *Dynamic-Struct* | R | 0.855 | 0.850 | 0.754 | 0.779 |
| | N | | 0.859 | | 0.724 |
| *Dynamic-Struct + atten* | R | 0.861 | 0.856 | 0.760 | 0.763 |
| | N | | 0.864 | | 0.756 |
| *Content Network* | R | 0.932 | 0.933 | 0.797 | 0.782 |
| | N | | 0.930 | | 0.810 |
| *Content + Structure* | R | 0.943 | 0.945 | 0.836 | 0.832 |
| | N | | 0.938 | | 0.839 |

(1) *Static-struct*: It is a graph classification method [21] which is used to classify the static structure in the final phase.

(2) *Dynamic-struct*: It only utilizes dynamic propagation structures without temporal attention mechanism.

(3) *Structure network (SN)*: It only utilizes dynamic structures with attention mechanism without the content information.

(4) *Content network (CN)*: It only utilizes the textual content without the dynamic structure information and temporal attention mechanism.

Experimental results of all components are shown in Table III. The *structure network* only uses the dynamic structural information without extra textual content information, which outperforms the feature-based approach in Table II. By incorporating the structure information, the performance of *CN + SN* is improved compared to *CN*, which demonstrates the dynamic structure information provides distinct evidence in rumor detection.

*3) Early Detection of Rumors*

To evaluate the performance of our proposed model in early detection of rumors, we compare with several other methods at different detection deadlines (see Fig. 3). We adopt the same detection deadline as [2, 3], which is the average of reporting time over rumors. From Fig. 3, we can see that our proposed model NM-DPS performs better than other methods at all different detection deadlines, demonstrating NM-DPS's effectiveness in the early detection of rumors.

## V. CONCLUSION

Detecting rumors in social media is an important task to avoid the wide spread of misinformation and minimize potential negative social consequences. Existing works on rumor detection only utilizes the limited static propagation structures rather than dynamic information. In this paper, we propose a temporal attention based model to extract features of dynamic propagation structures, and then integrates structural and textual information into a unified framework for effective rumor detection. Experimental results on two real social media datasets show that our model outperforms the state-of-the-art methods as well as the effectiveness of our model in utilizing the dynamic propagation structures.

## ACKNOWLEDGMENT

This work is supported in part by the National Key R&D Program of China under Grant No. 2016QY02D0305, the National Natural Science Foundation of China under Grants No. 71702181, No. 71621002 and No. 11832001, the National Science and Technology Major Project under Grant No. 2018ZX10201001 and the Early Career Development Award of SKLMCCS under Grant 20180208.